\documentclass[sn-mathphys-num]{sn-jnl}

\usepackage{graphicx}%
\usepackage{multirow}%
\usepackage{amsmath,amssymb,amsfonts}
\usepackage{amsthm}
\usepackage{mathrsfs}%
\usepackage[title]{appendix}%
\usepackage{xcolor}
\usepackage{textcomp}%
\usepackage{manyfoot}%
\usepackage{booktabs}%
\usepackage{algorithm}
\usepackage{algorithmic}
\usepackage{listings}%

\usepackage{colortbl}
\usepackage{booktabs}
\usepackage{multirow}
\usepackage{tcolorbox}
\usepackage{lipsum}
\usepackage{mdframed}

\theoremstyle{thmstyleone}%
\theoremstyle{thmstyletwo}%

\theoremstyle{thmstylethree}%

\usepackage{ifthen}
\usepackage[normalem]{ulem} 
\usepackage{xcolor}

\newboolean{showedits}
\setboolean{showedits}{true} 
\ifthenelse{\boolean{showedits}}
{
	\newcommand{\del}[1]{\textcolor{red}{\sout{#1}}} 
}{
	\newcommand{\del}[1]{} 
	
}

\newboolean{showcomments}
\setboolean{showcomments}{true} 
\newcommand{\id}[1]{$-$Id: scgPaper.tex 32478 2010-04-29 09:11:32Z oscar $-$}

\ifthenelse{\boolean{showcomments}}
{\newcommand{\nbc}[3]{
		{\colorbox{#3}{\bfseries\sffamily\scriptsize\textcolor{white}{#1}}}
		{\textcolor{#3}{$\blacktriangleright$#2$\blacktriangleleft$}}}
	}
{\newcommand{\nbc}[3]{}
	\renewcommand{\del}[1]{} 
	}

\definecolor{ibcolor}{rgb}{1.0,0.2,.4}
\definecolor{dsrcolor}{rgb}{0.5,0.6,0}
\definecolor{cfcolor}{rgb}{0,0.5,0.9}
\definecolor{oldcolor}{rgb}{0.2,0.2,0.2}
\definecolor{tdcolor}{rgb}{1.0,0,0}
\definecolor{oldcolor}{rgb}{0.5,0.5,0.5}
\definecolor{lycolor}{rgb}{0.3,0.3,0.8}

\begin{document}

\title[Article Title]{ABFS: Natural Robustness Testing for LLM-based NLP Software}


\author[1]{\fnm{Mingxuan} \sur{Xiao}}\email{xiaomx@hhu.edu.cn}

\author*[1,2]{\fnm{Yan} \sur{Xiao}}\email{xiaoyan.hhu@gmail.com}

\author[1]{\fnm{Shunhui} \sur{Ji}}\email{shunhuiji@hhu.edu.cn}
\author[1]{\fnm{Yunhe} \sur{Li}}\email{yunheli@hhu.edu.cn}

\author[2]{\fnm{Lei} \sur{Xue}}\email{xuelei3@mail.sysu.edu.cn}

\author*[1]{\fnm{Pengcheng} \sur{Zhang}}\email{pchzhang@hhu.edu.cn}

\affil*[1]{\orgdiv{Key Laboratory of Water Big Data Technology of Ministry of Water Resources \emph{and} the College of Computer Science and Software Engineering}, \orgname{Hohai University}, \orgaddress{\street{No. 8, Focheng West Road, Jiangning District}, \city{Nanjing}, \postcode{211100}, \country{China}}}

\affil*[2]{\orgdiv{School of Cyber Science and Technology}, \orgname{Shenzhen Campus of Sun Yat-sen University}, \orgaddress{\street{No.66 Gongchang Road, Xinhu Street, Guangming District}, \city{Shenzhen}, \postcode{528406}, \country{China}}}



\abstract{Owing to the exceptional performance of Large Language Models (LLMs) in Natural Language Processing (NLP) tasks, LLM-based NLP software has rapidly gained traction across various domains, such as financial analysis and content moderation. However, these applications frequently exhibit robustness deficiencies, where slight perturbations in input (prompt+example) may lead to erroneous outputs. Current robustness testing methods face two main limitations: (1) low testing effectiveness, limiting the applicability of LLM-based software in safety-critical scenarios, and (2) insufficient naturalness of test cases, reducing the practical value of testing outcomes.
To address these issues, this paper proposes ABFS, a straightforward yet effective automated testing method that, for the first time, treats the input prompts and examples as a unified whole for robustness testing. Specifically, ABFS formulates the testing process as a combinatorial optimization problem, employing \underline{B}est-\underline{F}irst \underline{S}earch to identify successful test cases within the perturbation space and designing a novel \underline{A}daptive control strategy to enhance test case naturalness.
We evaluate the robustness testing performance of ABFS on three datasets across five threat models. On Llama2-13b, the traditional StressTest achieves only a 13.273\% success rate, while ABFS attains a success rate of 98.064\%, supporting a more comprehensive robustness assessment before software deployment. Compared to baseline methods, ABFS introduces fewer modifications to the original input and consistently generates test cases with superior naturalness. Furthermore, test cases generated by ABFS exhibit stronger transferability and higher testing efficiency, significantly reducing testing costs.}

\keywords{Software testing, Test cases, NLP software, LLMs, Best-first search}



\maketitle

\section{Introduction}\label{sec1}

LLM-based NLP software, such as ChatGPT\footnote{\url{https://openai.com/chatgpt}}, New Bing\footnote{\url{https://www.bing.com}}, Gemini\footnote{\url{https://gemini.google.com}}, and Poe\footnote{\url{https://poe.com}}, has gained widespread attention and usage since its release~\cite{10.1145/3650212.3680343,bano2024large}. According to data analytics platforms like Coalition Technologies, as of July 2024, ChatGPT has approximately 180.5 million users, while New Bing has over 100 million daily active users, with combined software application downloads exceeding 26.6 million. In the field of NLP, users can provide prompts and examples as inputs to guide LLMs in completing various downstream tasks closely related to everyday life and industry needs~\cite{liu-etal-2024-stablept,zhu2023promptbench}. These tasks include public opinion monitoring~\cite{10.1007/978-3-031-61057-8_31}, financial analysis~\cite{10.1145/3650212.3680388}, and box office prediction~\cite{software3010004}. However, the high frequency of user interactions has inevitably intensified the negative impact of malicious inputs on the robustness of these software systems. Recent studies have shown that specific perturbations in the overall input (prompt+example) to LLM-based NLP software can mislead model outputs~\cite{10.1145/3691620.3695018,10.1145/3697012}, resulting in software misjudgments and functional anomalies, as demonstrated in the following scenario.

\begin{figure}[t]
 \centering
\includegraphics[width=0.9\linewidth]{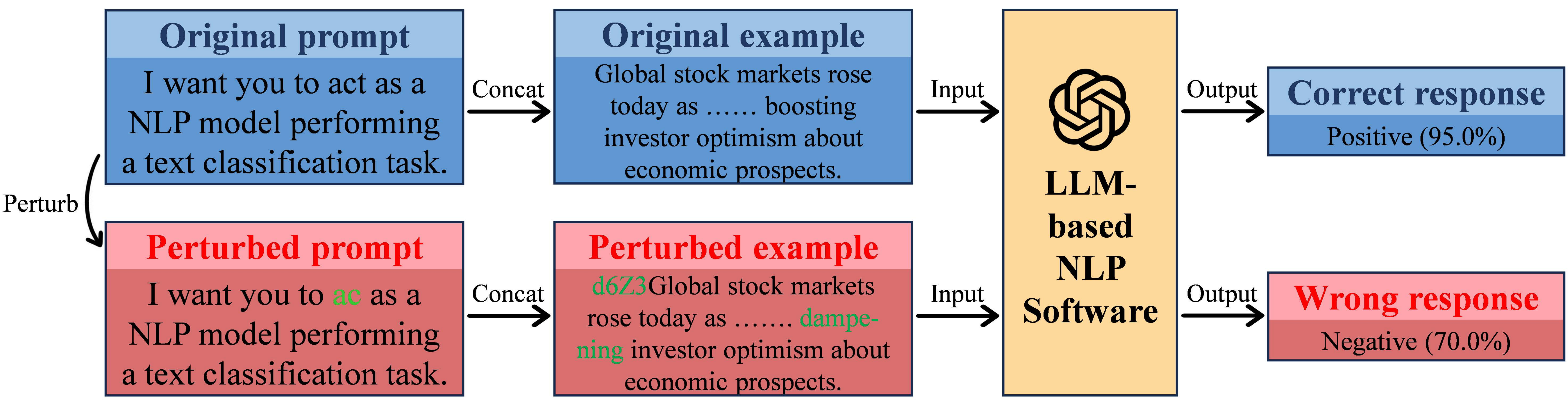}\\
 \caption{Slightly perturbed text (green) can mislead ChatGPT into judging the label of financial news from ``POSITIVE'' (with a confidence of 95\%) to ``NEGATIVE'' (with a confidence of 70\%).} 
 \label{Fig1}
 \vspace{-0.3cm}
\end{figure}

As shown in Fig.~\ref{Fig1}, the financial market exhibits a high sensitivity to news events, with significant news often triggering sharp market fluctuations. A timely grasp of sentiment orientation within news reports can help investors optimize their portfolios and improve returns. Since LLM-based software, such as ChatGPT, outperforms traditional DNNs~\cite{pangakis-wolken-2024-knowledge} and crowdsourced workers~\cite{gerosa2024can} in text classification, many companies are leaning toward LLM-based software for executing text classification tasks~\cite{jin2024llms}. However, when the overall input to the software—the text formed by appending the original prompt with financial news—is intentionally perturbed (e.g., changing ``boosting” to ``dampening”), ChatGPT o1-preview erroneously classifies the text as “Negative” instead of “Positive”. 
This misclassification, as documented in our reproducible repository\footnote{\url{https://github.com/lumos-xiao/ABFS}}, can lead investors to
take excessive preventive measures, such as unnecessary sell-offs or increased hedging, wasting resources and potentially causing market panic. With the rapid proliferation of LLM-based software, effective pre-release robustness testing has become a key focus of current research~\cite{10.1145/3655022,wu2024future}. On the other hand, the United States alone spends around \$48 billion annually on software testing~\cite{10.1145/3611643.3616327}, making it challenging to manually write numerous test cases for the LLM under test (i.e., the threat model). Inspired by fuzz testing~\cite{10449663}, we explore automated methods for robustness testing of LLM-based NLP software. By uncovering robustness flaws more efficiently, automated testing can bolster confidence in software quality while reducing testing costs and time.

We summarize the challenges faced by existing work as follows:

(1) \emph{There is an urgent need to improve the effectiveness of robustness testing for LLM-based NLP software.} Existing studies~\cite{10.5555/3666122.3668671, liu2023robustness, 10.1145/3691620.3695001, p2sql} tend to treat robustness testing for prompts and examples separately. However, LLM-based software outputs are based on the combined input (prompt + example) generated through calls to non-interpretable LLMs~\cite{zverev2024can}, making it fundamentally a black-box environment. To address this, our focus shifts to robustness concerning the overall input, utilizing black-box testing methods originally designed for DNN-based software as baselines for LLM-based software. Unfortunately, these baseline methods demonstrate significantly reduced effectiveness when applied to LLMs. For example, the well-known TextFooler~\cite{Jin_Jin_Zhou_Szolovits_2020} achieves 88.500\% and 87.500\% success rates on the AG’s News~\cite{zhang2015character} and MR~\cite{pang2005seeing} datasets when testing BERT~\cite{DBLP:conf/naacl/DevlinCLT19}, but only 38.363\% and 49.011\% when testing Llama2-70b~\cite{touvron2023llama}. This sharp decline highlights a critical gap in robustness testing for LLM-based NLP software. Low testing effectiveness results in undetected defects, which may lead to issues like data leakage, quality degradation, or increased maintenance costs post-release, severely impacting the reliability and security of the software.

(2) \emph{Generating test cases with subtle and natural variations from the original text is challenging.} Taking the financial analysis scenario illustrated in Fig.~\ref{Fig1} as an example, NLP software in real-world applications must handle various text inputs that include different wordings, synonyms, or slight grammatical changes. Generating test cases with minor yet natural modifications can better simulate real scenarios and assess the software's robustness when handling real-world data~\cite{tambon2022certify}. In Fig.~\ref{Fig1}, character deletions, special symbol insertions, and synonym replacements were used. While these techniques can mislead the software's decision-making, they contradict the core objective of robustness testing: assess the software's sensitivity to slight textual changes without altering the overall semantic meaning~\cite{chen-etal-2022-adversarial}. We observed that the currently best-performing baseline, PWWS~\cite{ren2019generating}, achieved a success rate of 49.302\% when testing Mistral-7b~\cite{jiang2023mistral} on the MR dataset. However, the original text modification rate (change rate) and text naturalness (perplexity) results were 6.145\% and 78.205, respectively. These results suggest that while the approach generates effective test cases, it fails to balance subtlety and naturalness optimally. Generating test cases through subtle textual perturbations effectively reveals performance flaws of the threat model under boundary conditions, providing valuable insights for software updates and iterations.

Leveraging the black-box interaction between NLP software and LLMs, we propose an automated robustness testing method named ABFS, addressing a significant research gap in generating natural test cases for LLM-based NLP software.
ABFS employs Best-First Search (BFS)~\cite{4082128} to navigate the feature space of the threat model in search of successful test cases, selecting the optimal candidate case at each step for further exploration. This selection strategy ensures that each modification is currently optimal, guiding the search toward successful tests while maintaining textual naturalness. Although BFS uses a locally optimal strategy, it avoids local optima by managing nodes to be explored through a priority queue, thereby increasing the probability of successful tests.
To further enhance BFS, we incorporate an adaptive strategy that combines the current state with historical performance to evaluate each word's impact on software robustness precisely. This refinement allows ABFS to perturb fewer words, thereby improving the naturalness of the generated test cases.

In this study, we experiment with three datasets related to sentiment monitoring, financial analysis, and box office prediction to assess ABFS's robustness testing capabilities across five threat models of different parameter scales. Results show that existing methods fail to test LLM-based NLP software effectively; for example, StressTest~\cite{naik2018stress} achieves an average success rate of 13.460\% on AG's News, whereas ABFS reaches 64.330\%. This indicates that ABFS uncovers more robustness defects during the software testing phase. We also evaluate the perturbation extent and naturalness of ABFS-generated test cases using change rate and perplexity. 
Under 15 experimental settings, ABFS achieves a change rate of 1.009\% and a perplexity of 49.534, demonstrating that ABFS introduces smaller perturbations to the original text and generates more natural test cases.

The contributions of this paper include: 
\begin{itemize}
\item Applying BFS to the robustness testing of LLM-based NLP software for the first time, we focus on its unique input structure (prompt + example) and the complex interaction characteristics inherent to LLMs. By utilizing the priority queue structure of BFS, global optimization is achieved in the high-dimensional natural language feature space of LLMs, significantly enhancing the effectiveness of robustness testing.

\item Introducing an adaptive control strategy to improve BFS, we propose ABFS, a robustness testing method designed for LLM-based NLP software. Conventional backtracking mechanisms tend to repeat ineffective test cases~\cite{10.1145/3588695}, but ABFS optimizes perturbation order by updating based on each word's historical score, enabling greater misdirection of software outputs with fewer perturbation steps. This further enhances the naturalness of the generated test cases.

\item Conducting extensive experiments on three datasets, comparing ABFS against five baselines and across five LLMs of varying parameter sizes. The results demonstrate that test cases generated by ABFS achieve higher testing success rates, superior change rates, and better perplexity scores. Additionally, we conduct tests on efficiency, transferability, and perform ablation studies to evaluate ABFS's performance further. The data and code are available in our reproducible repository\footnote{\url{https://github.com/lumos-xiao/ABFS}}.
\end{itemize}

The remainder of this paper is organized as follows: Section~\ref{sec2} introduces BFS and the basic concepts of robustness testing for LLM-based software. In Section~\ref{sec3}, we propose a black-box testing method, ABFS, to generate natural test cases. In Section~\ref{sec4}, we validate ABFS using five threat models, five baselines, and three datasets. Experimental results and the effectiveness of the ABFS testing strategy are presented in Section~\ref{sec5}. Section~\ref{sec6} discusses the threats to validity faced by this study. Section~\ref{sec7} reviews existing work on robustness testing for NLP software. Finally, Section~\ref{sec8} concludes this paper.

\section{Preliminary}\label{sec2}
\subsection{Best-first search (BFS)}
BFS is a class of heuristic search algorithms that uses a heuristic function $H(e_n)$ to guide the search process, prioritizing the expansion of nodes deemed closest to the goal. This algorithm was first introduced by Hart et al.~\cite{4082128} in 1968 for solving path-planning problems. The core concept of BFS is to use $H(e_n)$ to estimate the cost from each node to the target node, with the consistency of the heuristic function required to satisfy the following inequality:
\begin{equation}
H({e_n}) \le c({e_n},{e_m}) + H({e_m})
\end{equation}
where $c({e_n},{e_m})$ represents the actual cost from node $e_n$ to its neighboring node $e_m$. By expanding nodes in order of increasing cost, BFS ensures the discovery of an optimal solution. The algorithm maintains a priority queue to store nodes pending expansion, each time selecting the node with the smallest heuristic value for expansion. The specific steps are as follows:

(1) Initialization: Insert the starting node $e_0$ into the priority queue with an initial heuristic value $H(e_0)$, and initialize an empty set to store expanded nodes.

(2) Node Selection: Retrieve the node $e_n$ with the smallest heuristic value from the priority queue. If this node is the target, the algorithm terminates and returns the path.

(3) Node Expansion: If $e_n$ is not the target node, mark it as expanded and process all its neighboring nodes. If each neighbor $e_m$ has yet to be expanded, insert it into the priority queue and calculate its heuristic value $H(e_m)$.

(4) Repeat Steps: Repeat the node selection and expansion steps until the target node is found or the priority queue is empty.

BFS is commonly applied in large search spaces where solutions are not immediately apparent, such as search-based software engineering~\cite{9987661,1702110}, processor scheduling~\cite{KISHIMOTO2013222}, and integrated circuit hardware verification~\cite{1512370}. 
The algorithm's adaptability makes it particularly effective for problems involving incomplete information or high complexity, achieved through a straightforward heuristic-driven search process.
However, BFS has limitations. Its performance depends heavily on the stability of the heuristic function, and it can encounter local optima that hinder progress. To mitigate these issues, recent research incorporates enhancements such as local search restarts or backtracking mechanisms, 
which is prone to redundant perturbations~\cite{10.1145/3588695}. In contrast, ABFS proposed in this study adaptively considers the historical information of nodes during the search process, ensuring that each perturbation moves towards the global optima, thereby effectively improving optimization performance.

\subsection{Problem definition}
LLM-based NLP software has been deployed in various complex scenarios,
and complex applications demand high levels of robustness from LLMs. In safety-critical contexts, such as command interpretation in autonomous driving or transaction instructions in financial systems, robustness testing ensures that software can handle edge cases and abnormal inputs without catastrophic failure~\cite{p2sql,bano2024large}. It is worth noting that LLMs function within a high-dimensional feature space and are susceptible to the overall input~\cite{10.1145/3691620.3695001, zverev2024can}. This high dimensionality and sensitivity make LLM-based software vulnerable to input perturbations, resulting in inconsistent output. Compared to traditional DNNs, LLMs often have significantly more parameters and a more complex architecture~\cite{zhu2023promptbench}, increasing the difficulty of predicting and controlling their behavior. Robustness testing can identify vulnerabilities within the expansive feature space of LLMs, allowing timely detection and correction of robustness defects to prevent costly maintenance in production environments.

The IEEE defines robustness in software engineering as ``the degree to which a system, product, or component performs its specified functions under specified conditions for a specified period."~\cite{iso2017iso}. This study focuses on robustness in text classification tasks, as classification serves as the foundation for many other downstream NLP applications~\cite{tung2024automated}. 
Strong classification performance provides insight into the decision paths of the threat model, enabling developers to monitor and refine its behavior.
Given the LLM $\mathcal{F}$ as the decision-making component used by LLM-based NLP software, its training dataset is $(x, y) \sim \mathcal{D}$, where $x$ represents the example and $y$ is the ground-truth label, we create the software input $I=[P;x]$ by concatenating a preset prompt $P$ with $x$. The robustness of the software can be evaluated by assessing $\mathcal{F}$'s predictive performance on test cases $\left(I^{\prime}, y^{\prime}\right) \sim \mathcal{D}^{\prime}\neq\mathcal{D}$: the greater the decline in $\mathcal{F}$'s predictive accuracy on $\mathcal{D}^{\prime}$ compared to its original accuracy on $\mathcal{D}$, the fewer test cases $\mathcal{F}$ correctly classifies, indicating weaker robustness of the software.

In robustness testing, a test case consists of a perturbed input $I'$ and an expected result representing the original input's label. ABFS is designed to automate the generation of natural adversarial test cases, a concept first introduced by Szegedy et al.~\cite{DBLP:journals/corr/SzegedyZSBEGF13}. Adversarial test cases are created by adding slight perturbations $\delta$ to the original inputs, which impact the threat model $\mathcal{F}$ but remain imperceptible to humans~\cite{10.1007/s10515-024-00470-9}. This results in an adversarial test case $I_{adv}$, which induces a response $\mathcal{F}(I_{adv})$ different from the original output $\mathcal{F}(I_{ori})$. Given the original input $I_{ori}=[P;x]$ and the threat model $\mathcal{F}$, robustness testing using adversarial test cases can be represented as:
\begin{equation}
\begin{aligned}
& \underset{I_{a d v} \in C\left(I_{o r i}\right)}{\textit{arg\,min} }\left\|I_{o r i}, I_{a d v}\right\| \\
& \text { s.t. } \mathcal{F}\left(I_{o r i}\right) \neq \mathcal{F}\left(I_{a d v}\right)
\end{aligned}
\label{eq2}
\end{equation}
where $\|t_1, t_2\|$ denotes the difference between two text segments $t_1$ and $t_2$. ABFS represents this difference using a change rate to generate natural test cases. The constraint $C$ ensures the quality of textual modifications, which in ABFS includes filtering stop-words and limiting the maximum change rate to maintain naturalness.

\section{Methodology}\label{sec3}
\begin{figure}[t]
 \centering
\includegraphics[width=0.8\linewidth]{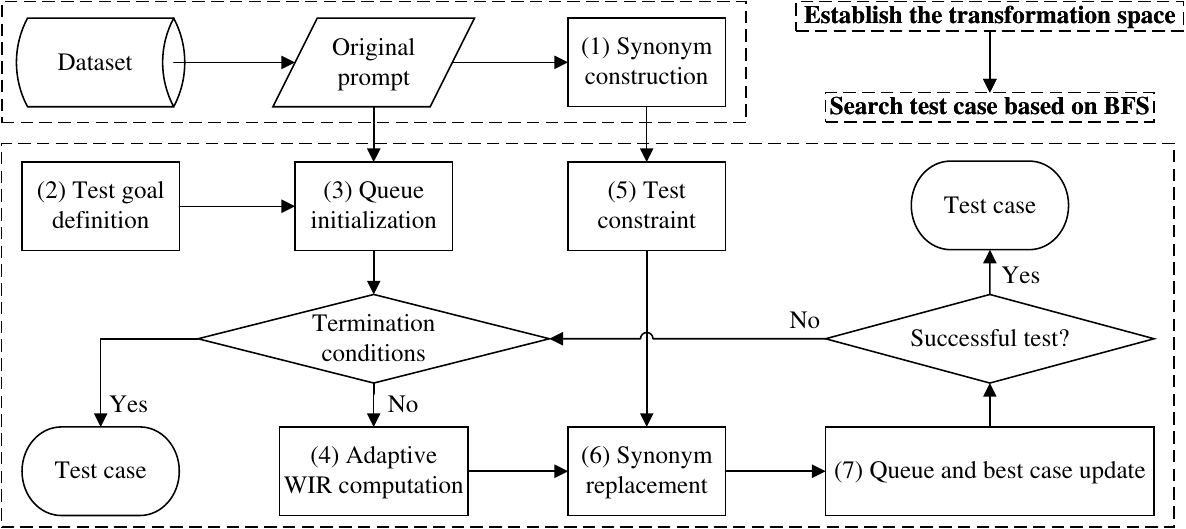}\\
 \caption{Overview of ABFS.} 
 \label{Fig2}
\end{figure}

Fig.~\ref{Fig2} provides an overview of the proposed ABFS, designed to automate the generation of natural test cases for LLM-based NLP software. First, text examples from the test dataset are concatenated with a given prompt to form the original input $I_{ori}$. Then, (1) use WordNet~\cite{10.1145/219717.219748} to build a synonym vocabulary for all words in $I_{ori}$. (2) Formalize the goal function of robustness testing for BFS. (3) Establish a priority queue using the combination of $I_{ori}$ and its heuristic value as the starting and optimal node. Next, the iterative deepening search is performed, checking termination conditions on the current queue. If the queue is empty or the maximum query limit for the threat model has been reached, the optimal node is output as the test case; otherwise, (4) an adaptive strategy calculates the importance of each word by index to determine the order of perturbation. Combined with (5) testing constraints, the most important word undergoes (6) synonym replacement, generating a set of candidate subnodes. (7) Insert subnodes with heuristic values higher than the current optimal node into the priority queue and update the optimal node with the best-performing new node. If the new optimal node successfully misleads the threat model's prediction, the test is considered successful, and the optimal node is output as the test case; otherwise, the iterative search continues on the current queue.

The description of ABFS is divided into two sections: establishing the transformation space, which involves defining the permissible perturbations, and searching for test cases based on BFS, where the algorithm navigates the transformation space to identify optimal adversarial test cases.
\subsection{Establishing the transformation space}
\begin{figure}[t]
 \centering
\includegraphics[width=0.6\linewidth]{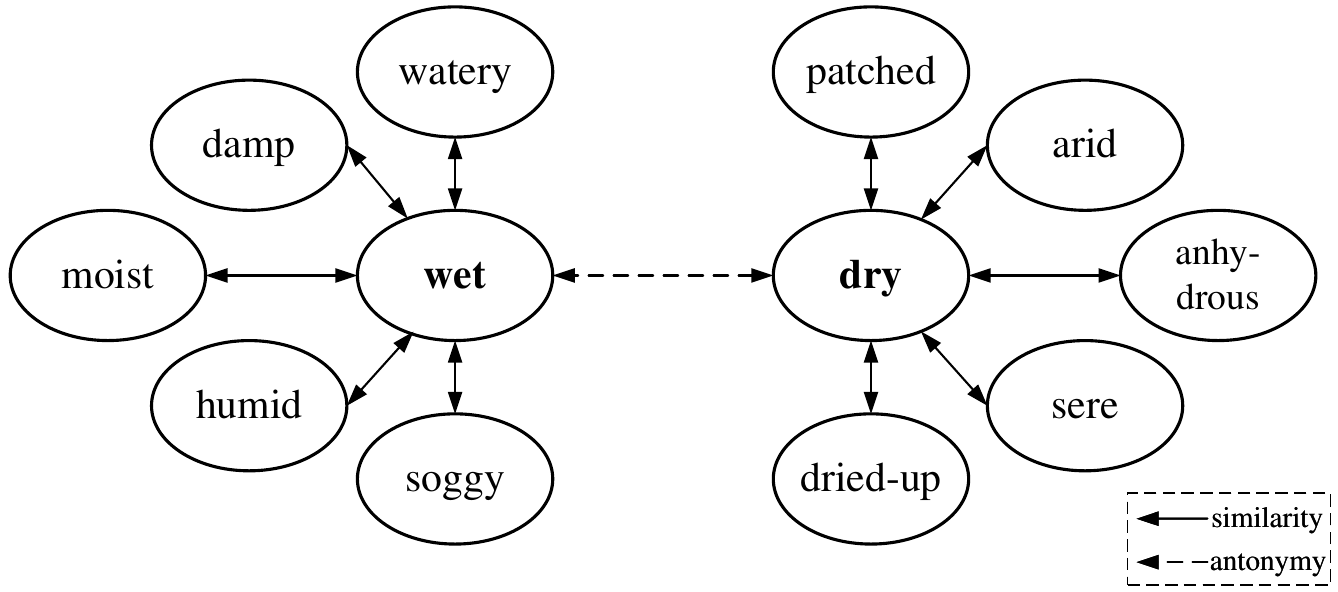}\\
 \caption{Bipolar adjective structure of WordNet.} 
 \label{Fig3}
\end{figure}

To generate adversarial test cases, it is essential to establish a transformation space through subtle perturbations for ABFS to conduct its search. In the context of robustness testing for LLM-based NLP software, the opaque relationship between LLMs, examples, and prompts necessitates treating them as a combined entity for analysis. Each time, we sample a text instance from the test data and concatenate it with a prompt relevant to the downstream task, forming $I_{ori}$, which will be transformed. Synonym replacement is a common perturbation technique; ABFS leverages the English lexical database WordNet to build a synonym vocabulary~\cite{10.1145/219717.219748}, grouping vocabulary according to semantic relationships, particularly organizing synonyms into ``synsets". Each synset represents a specific semantic unit, making WordNet an ideal tool for synonym replacement, especially when creating natural test cases that do not alter semantics. For example, in Fig.~\ref{Fig3}, (``wet", ``dry") is an antonym pair represented by a dashed line, while (``wet", ``damp") or (``dry", ``arid") are synonym pairs represented by solid lines.

ABFS follows these steps to establish the transformation space using WordNet:

Part-of-Speech Tagging: Before synonym replacement, each word in $I_{ori}$ is tagged for its part of speech. WordNet organizes synonyms by part of speech (such as nouns, verbs, adjectives, and adverbs), and accurate tagging helps retrieve the correct synonyms.

Synonym Retrieval: Based on part-of-speech tagging, ABFS retrieves the synsets of target words from WordNet. Each synset represents a specific semantic concept and includes all associated synonyms.

Semantic Filtering: Not all synonyms can maintain semantic consistency in context, so further filtering is required. ABFS calculates the cosine similarity between each synonym and the original word, retaining only the top $k_1$ synonyms with the highest similarity. The filtered synonyms for each word form the transformation space for ABFS's search.

The core advantage of using WordNet to establish a synonym vocabulary is its ability to generate highly natural and semantically consistent test cases. These cases successfully mislead the threat model's decision-making through subtle synonym replacements that do not alter the original semantics, providing an in-depth robustness assessment for LLM-based software.

\subsection{Searching test cases based on BFS}
Currently, there is a lack of work focused on testing the robustness of LLM-based NLP software concerning the overall input. Our experimental results underscore the main challenge in robustness testing for LLM-based software: the effectiveness of applying existing DNN-based software testing methods to LLM-based software is shallow (RQ1). To improve the practical utility of robustness testing, we introduce the novel ABFS for LLM-based software testing, which searches within the transformation space to find test cases that meet the goal function.
\subsubsection{Test goal definition}
To transform automated testing into a combinatorial optimization problem suitable for BFS, we first define the test goal function according to $Eq.\ref{eq2}$. Given a dataset $\mathcal{D}={\{ ({x_i},{y_i})\} _{i \in [N]}}$ and an initial prompt $P$, robustness testing concerning input aims to mislead the threat model $\mathcal{F}$ by applying a perturbation $\delta$ within the given constraint $C$ to the concatenated text $[P;x]$ derived from the example and prompt:
\begin{equation}
{\textit{arg\,min} _{\delta  \in C}}{E_{(x;y) \in \mathcal{D}}}\mathcal{L}[\mathcal{F}([P;x] + \delta ),y]
\end{equation}
As shown in Fig.~\ref{Fig1}, $P$ might be ``I want you to act as a natural language processing model performing a text classification task. My first text is:
”, and $x$ could be ``Global stock markets rose today ......, boosting investor optimism about economic prospects.
”. Here, $\delta$ represents a perturbation applied to the entire “prompt + example” input, and $\mathcal{L}$ is the confidence corresponding to the label output by the threat model. Confidence quantifies $\mathcal{F}$’s certainty in its output result, and the lower the confidence in the original label $y$ after perturbation, the closer the current input is to flipping the threat model’s decision label, indicating a near-successful test case. This definition is analogous to adversarial test case generation in DNN-based software testing, and we extend this concept to the robustness testing of LLM-based software.
\subsubsection{Adaptive best-first search}
\begin{figure}[t]
 \centering
 \vspace{-0.3cm}
\includegraphics[width=1\linewidth]{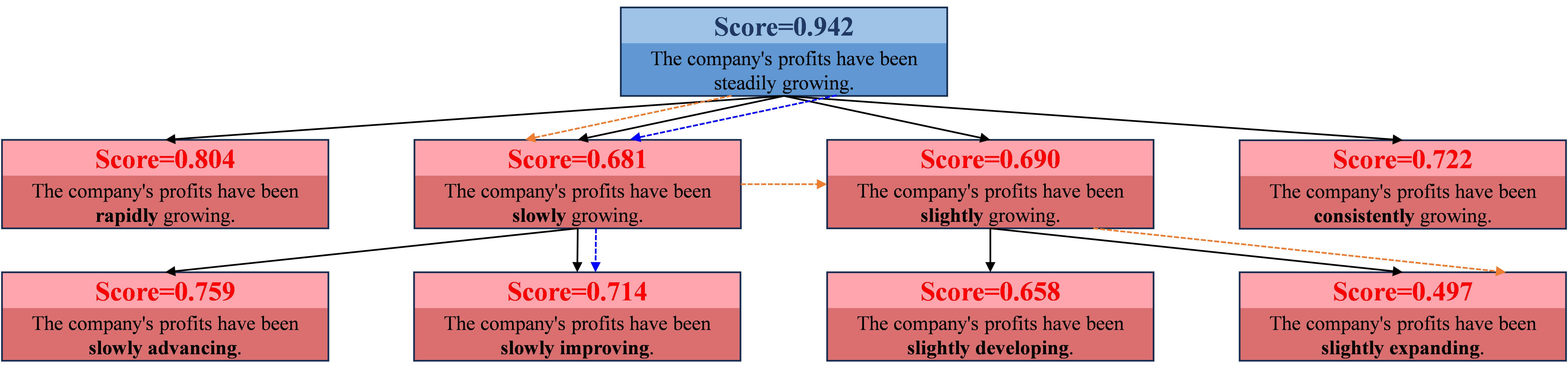}\\
 \caption{An example of searching for successful test cases in the transformation space, where black arrows represent synonym replacement operations, blue and orange dashed arrows indicate the paths taken by the greedy search and BFS, respectively. ``Score" represents the confidence score of the threat model for the ground truth label of the input, with lower scores indicating a more effective search strategy.} 
 \label{Fig4}
 \vspace{-0.4cm}
\end{figure}
In NLP test case generation, the search space can be represented as a directed acyclic graph, where nodes represent modified text, and edges denote synonym replacement operations. The testing goal aims to reduce the threat model's confidence score $\mathcal{L}[\mathcal{F}(I),y]$ on the input text. In greedy search, only the neighboring node with the lowest confidence score is selected at each step: ${e_{t + 1}} = \mathop {\textit{arg\,min}}\limits_{e' \in N({e_t})} H(e')$. Although this method is simple and fast, it tends to explore only nodes in the local open set, making it susceptible to local optima and often unable to reach the globally optimal solution in the closed set. As shown in Fig.~\ref{Fig4}, greedy search stalls at a node with a confidence score of 0.714, unable to reduce it further. BFS also uses confidence as the goal function. However, by maintaining a priority queue $Q$, it selects the optimal node from the entire queue at each step rather than limiting itself to current neighboring nodes: ${e_{t + 1}} = \mathop {\textit{arg\,min}}\limits_{e' \in Q} H(e')$. This global exploration strategy gradually transforms open-set into closed-set nodes, avoiding local optima and ultimately finding a node with a lower confidence score, such as ``Score=0.497", as illustrated by the brown path in Fig.~\ref{Fig4}. Therefore, BFS, compared to greedy search, more effectively generates test cases. Algorithm~\ref{algorithm1} outlines the search process of ABFS.
\begin{algorithm}[t]
    \footnotesize
    \caption{Search Process in ABFS}
    \label{algorithm1}
    \begin{algorithmic}[1]
	\REQUIRE $x_{ori}$: original text input, $P$: original prompt input, $\mathcal{F}$: threat model, $max\_query$: maximum query number.
	\ENSURE $I_{adv}$: adversarial test case.
        \STATE $queue$$\leftarrow$PriorityQueue($\emptyset$);
        \STATE $best\_case$$\leftarrow$Concat($P$, $x_{ori}$);
        \STATE $queue\leftarrow queue \cup \{(\mathcal{F}(best\_case), best\_case)\}$;
        \WHILE{not exceed $max\_query$ $\lor$ $queue$$\neq\emptyset$}
        \STATE $I_{cur}$$\leftarrow$Dequeue($queue$);
        \STATE $index\_score$$\leftarrow$$WIR(I_{cur})$ via $Eq.\ref{eq4}$;
        \FOR{$word\in I_{cur}$}
        \STATE add $\Delta(\mathcal{F}(I_{cur}),index\_score[word])$ into $index\_history[word]$;
        \STATE $score_{\delta}$$\leftarrow$Average($index\_history[word]$) via $Eq.\ref{eq5}$;
        \STATE $index\_score[word]$+=$score_{\delta}$;
        \ENDFOR
        \STATE $index\_order$$\leftarrow$argsort($index\_score$,order=descend);
        \FOR{$idx \in index\_order$}
        \STATE $nodes\leftarrow$Transform($I_{cur},idx$) via $Eq.\ref{eq6}$;
        \FOR{$node \in nodes$}
            \IF{$\mathcal{F}(node) < \mathcal{F}(best\_case)$}
                \STATE $queue \leftarrow queue \cup \{(\mathcal{F}(node),node)\}$;
            \ENDIF
        \ENDFOR
        \STATE $score_{Q}$$\leftarrow$[score \textbf{for} score $\in$ $\mathcal{F}$($queue$)];
        \STATE $best\_case$$\leftarrow$argsort($score_{Q}$,order=ascend)[0];
        \IF{$I_{adv}$ in $nodes$}
        \RETURN $I_{adv}$$\leftarrow$$best\_case$;
        \ENDIF
        \ENDFOR
        \ENDWHILE        
        \RETURN $I_{adv}$$\leftarrow$$best\_case$;
 \end{algorithmic}
\end{algorithm}

After setting the reduction of the threat model's confidence as the testing goal, the queue initialization (Lines 1-3) begins by establishing an empty priority queue. A priority queue is an abstract data structure that supports two fundamental operations: the Enqueue operation to insert elements and the Dequeue operation to remove elements. Each element $e_i$ has an associated priority $p(e_i)$ in a given min-priority queue. Dequeue removes the element with the highest priority: ${e_{\textit{min} }} = \mathop {\textit{arg\,min} }\limits_{{e_i} \in Q} p({e_i})$. ABFS implements the priority queue using a min-heap, where elements with lower priority values are dequeued first. Enqueue and Dequeue have a time complexity of $O(\log n)$, where $n$ is the number of elements in the queue. In robustness testing, the commonly used greedy search generally requires $m$ steps, with each step iterating over $n$ candidate solutions, resulting in a time complexity of $O(m \cdot n)$. In NLP software, the search space grows exponentially with the number of words, meaning $m$ is significantly larger than $n$, which gives the priority queue a substantial performance advantage. Queue initialization also includes concatenating the original example $x_{ori}$ sampled from the dataset with the NLP task-oriented prompt $P$ to form the original input. This original input is then inserted into the priority queue as the starting and optimal node, with its priority set as the confidence score of the threat model on the original input.

During the iterative search phase (Lines 4-26), the element with the highest priority is dequeued as the current input $I_{cur}$ for the software (Line 5), ABFS determines the perturbation order on $I_{cur}$ based on adaptive Word Importance Ranking (WIR)~\cite{Jin_Jin_Zhou_Szolovits_2020}. ABFS calculates the importance of each word in $I_{cur}$ by sequentially deleting individual words (Line 6), for a given input text ${I_{cur}} = \{ {w_1},{w_2},...,{w_i},...,{w_n}\}$ and the text $I_{cur}^{{w_i}} = \{ {w_1},{w_2},...,{w_{i - 1}},{w_{i + 1}},...,{w_n}\}$ obtained after deleting a word $w_i$, the word importance score of $w_i$ can be represented as:
\begin{equation}
\operatorname{WIR}\left(I_{c u r}, w_i\right)=\left\{\begin{array}{c}
\mathcal{L}\left[\mathcal{F}\left(I_{c u r}\right), y\right]-\mathcal{L}\left[\mathcal{F}\left(I_{c u r}^{w_i}\right), y_{w_i}\right], \text { if } y=y_{w_i} \\
\mathcal{L}\left[\mathcal{F}\left(I_{c u r}\right), y\right]+\mathcal{L}\left[\mathcal{F}\left(I_{c u r}^{w_i}\right), y_{w_i}\right]-\mathcal{L}\left[\mathcal{F}\left(I_{c u r}^{w_{i}}\right), y\right]\\
-\mathcal{L}\left[\mathcal{F}\left(I_{c u r}\right), y_{w_i}\right], \text { if } y \neq y_{w_i}
\end{array}\right.
\label{eq4}
\end{equation}
where $y$ represents the prediction label of the threat model for $I_{cur}$, and $y_{wi}$ represents the prediction label of the threat model for $I_{c u r}^{w_{i}}$. This method effectively tests the local sensitivity of LLM-based software to input. Important words carry critical information highly correlated with classification labels, and removing these words leads to misclassification, indicating their importance. By sequentially deleting each word in $I_{cur}$ and recording prediction changes after each deletion, an index list reflecting word importance ranking can be obtained, providing a basis for subsequent synonym replacement.

ABFS dynamically adjusts the importance score $S$ for each word index by recording the historical changes in word importance (Lines 7-11). Given the historical importance changes for a word $w_i$ as $\{ \Delta {S_{i1}},\Delta {S_{i2}},...,\Delta {S_{in}}\}$, where the score change $\Delta {S_i} = S({I_{cur}}) - S(I_{cur}^{{w_i}})$, ABFS adaptively calculates the average score change based on the most recent $k_2$ records:
\begin{equation}
\overline {\Delta {S_i}}  = \frac{1}{k}\sum\nolimits_{j = 1}^k {{\alpha _j}\Delta {S_{ij}}}
\label{eq5}
\end{equation}
where $\alpha _j$ is a hyperparameter, the adjusted word importance can be represented as ${S_i}^\prime  = {S_i} + \overline {\Delta {S_i}}$. The adaptive WIR ensures that the importance score depends not only on the current single perturbation result but also considers past performance, thus more accurately reflecting each word's importance. This enables the effective identification and selection of words that have the most significant impact on software robustness, thereby enhancing the naturalness of the perturbed text. On the other hand, by averaging the importance changes over the most recent $k_2$ instances, the fluctuation of importance scores is smoothed, preventing drastic changes due to anomalous single perturbation results and improving the stability of WIR. Sorting the index list in descending order of the adjusted word importance scores yields the sequence of words for ABFS to apply perturbation transformations (Line 12).

ABFS applies stop-word filtering and a maximum change rate as constraint conditions in the test case generation process. Stop words are words that frequently appear in text but contribute little to the text's semantics, such as functional words like ``of", ``is", ``to", etc. In practical applications, modifying stop words is easily detectable by humans and grammar-checking tools, so stop-word filtering helps prevent the generation of evident, unnatural perturbations. The maximum change rate refers to the upper limit on the proportion of words that can be modified by the testing method. If too many words are altered in the text, the generated test case will likely lose semantic consistency with the original sentence. Based on the reordered word index list and test constraint $C$, ABFS sequentially transforms each word $w_p$ to be perturbed (Line 14):
\begin{equation}
I_{cur}^{\prime}= \begin{cases}T\left(I_{cur}, w_p, w_p^{\prime}\right), & \text { if } w_p^{\prime} \in \operatorname{Syn}\left(w_p\right) \text { and } C\left(I_{cur}^{\prime}\right)=1 \\ I_{cur}, & \text { if } C\left(I_{cur}^{\prime}\right)=0\end{cases}
\label{eq6}
\end{equation}
where $T$ represents synonym replacement, and all transformed texts constitute the open-set nodes in ABFS's local search for this iteration. Nodes from the open set that achieve a greater reduction in confidence compared to the current best test case are selected and added to the closed set maintained by ABFS's priority queue (Lines 15-19). The best test case is updated based on the confidence scores output by the threat model for the closed-set nodes (Lines 20-21). Finally, ABFS checks if any of the nodes generated in this iteration can mislead the threat model's prediction label; if such a case exists, it is output as the test case; otherwise, the next round of perturbation is performed (Lines 22-24). To balance testing effectiveness with computational resources, when the maximum query limit is reached or the priority queue is empty, the best test case in the queue is output regardless of whether a successful test case has been found (Line 27). Even if testing fails, the best test case represents the historically optimal example generated by ABFS, capable of exerting the maximum perturbation on the LLM-based software, providing valuable insights for future robustness studies.

Overall, ABFS introduces a BFS strategy to provide a simple and effective combinatorial optimization method for robustness testing of LLM-based NLP software inputs, overcoming the limitations of current studies in handling input. By employing an adaptive control strategy, the assessment of word importance is not only based on the current perturbation results but also considers historical performance, thereby optimizing the perturbation order and further enhancing the effectiveness and naturalness of the test cases.
\section{Experiment setup}\label{sec4}
To verify the testing effectiveness of ABFS on LLM-based NLP software, we conducted a series of experiments on three text classification datasets and five threat models. All experiments were performed on an Ubuntu 22.04.1 LTS system equipped with two 32-core Intel(R) Xeon(R) Platinum 8358 CPUs at 2.60GHz, four NVIDIA A100 Tensor Core GPUs, and 1TB of physical memory. Each experiment was repeated three times, and the results for each metric were averaged. Similar to previous studies, 1,000 examples were randomly selected for testing in each experiment for each threat model. Therefore, the scale of these experiments is sufficient to cover various input data types, ensuring the representativeness and credibility of the experimental results. ABFS tests software based on the BFS algorithm and includes only three hyperparameters. After parameter tuning, we set $k_1$=25, $k_2$=5, and $\alpha _j$=1.
\subsection{Datasets}
Text content generated by different domains and users varies greatly, encompassing diverse grammatical structures, language conventions, and specialized terminology. Robustness testing ensures the software performs well even when confronted with diverse inputs. In this study, we select three datasets to represent classification tasks for financial texts, news texts, and movie review texts, covering different text lengths and binary/multi-class text classification tasks:

\textit{Financial Phrasebank (FP)}\footnote{\url{https://huggingface.co/datasets/financial_phrasebank}}~\cite{malo2014good}. A dataset of English news articles from all companies listed on OMX Helsinki. From this news database, 10,000 articles are randomly selected to achieve a representation of both small and large companies, firms across different industries, and reports from various news sources. The dataset is annotated by 16 individuals with sufficient background knowledge of financial markets. A sample of 4,840 sentences is randomly selected to represent the entire database.

\textit{AG's News}\footnote{\url{https://s3.amazonaws.com/fast-ai-nlp/ag\_news\_csv.tgz}}~\cite{zhang2015character}. This dataset quotes 496,835 news articles from more than 2,000 news sources in the 4 classes of AG's News Corpus (World, Sports, Business, and Science/Technology) in the title and description fields. We concatenate the title and description fields of each news article and use the dataset organized by kaggle\footnote{\url{https://www.kaggle.com/amananandrai/ag-news-classification-dataset}}. Each class contains 30,000 train examples and 1,900 test examples.

\textit{MR}\footnote{\url{https://huggingface.co/datasets/rotten_tomatoes}}~\cite{pang2005seeing}. A dataset widely used in sentiment polarity classification of movie reviews. This dataset includes 10,662 review summaries from Rotten Tomatoes, written by professional film critics. Each review text is accompanied by a binary label indicating the review's sentiment orientation (positive or negative). These labels are assigned based on the star ratings contained within the reviews.

\subsection{Threat models}
Five open-source LLMs are chosen as threat models to assess the ABFS testing performance on various LLM-based NLP software. This selection is crucial because security-sensitive sectors like finance and healthcare demand comprehensive security audits and proof of compliance. Open-source LLMs offer transparency that aids organizations in meeting these requirements more effectively than closed-source LLMs. Additionally, choosing open-source LLMs improves the verifiability and reproducibility of studies. Based on the difference in the number of parameters, we select five threat models on Hugging Face, including Mistral-7b-Instruct-v0.2~\cite{jiang2023mistral}, Llama-2-13b-chat~\cite{touvron2023llama}, Internlm2-chat-20b~\cite{cai2024internlm2}, Yi-34b-Chat~\cite{young2024yi}, and Llama-2-70b-chat~\cite{touvron2023llama}, to verify the generalizability of the test methods and the effect of model complexity on robustness.
\subsection{Baselines}
Existing work has separated robustness testing of software concerning prompts and examples, and currently, there is a lack of comparative methods for robustness testing of overall input. We observe that existing testing methods designed for DNN-based NLP software, which treat examples as input, are procedurally similar to testing LLM-based software. In our preliminary evaluation, we found that methods employing complex meta-heuristic search algorithms, such as genetic algorithm~\cite{ALETI2015343} and particle swarm optimization~\cite{10298415}, incur significant computational costs (exceeding 500 minutes each), which contradicts the goal of automated testing to improve development efficiency and software maintainability. Therefore, we adapt five DNN-based software testing methods guided by greedy strategy to the robustness testing context of LLM-based NLP software.
Most baselines are proposed before 2022 because current robustness testing for DNN-based NLP software primarily focuses on hard-label scenarios, where perturbations are evaluated solely based on output labels. This setting increases the adversarial difficulty for attackers but makes it challenging for software testers to capture more fine-grained software behaviors. In contrast, both the baselines used in this paper and ABFS adopt a confidence-guided soft-label robustness testing approach, enabling a more detailed analysis of software behavior under different inputs and significantly enhancing the comprehensiveness of the testing process.
Specifically, these methods include:

(1) \emph{CheckList} proposed by Ribeiro et al.~\cite{2021Beyond}: inspired by principles of behavioral testing in software engineering, CheckList guides users in what to test by providing a list of linguistic capabilities. To break down potential capability failures into specific behaviors, CheckList introduces different test types and then implements multiple abstractions 
to generate adversarial test cases.

(2) \emph{StressTest} proposed by Naik et al.~\cite{naik2018stress}: through the idea of ``stress test'', software is tested beyond its normal operational capabilities to detect weaknesses. Specific adversarial test case construction involves using heuristic rules with external sources of knowledge for competence tests, propositional logic frameworks for distraction tests, and random perturbations for noise tests.

(3) \emph{PWWS} proposed by Ren et al.~\cite{ren2019generating}: based on the synonym replacement strategy, a new word replacement order determined by the significance of the word and the classification probability is introduced, which belongs to the greedy search method.

(4) \emph{TextBugger} proposed by Li et al.~\cite{li2019textbugger}: first finds the important statements based on the degree of description of the facts; then uses a scoring function to determine the importance of each word to the classification result and ranks the words based on the scores; and finally use the proposed bug selection algorithm to change the selected words.

(5) \emph{TextFooler} proposed by Jin et al.~\cite{Jin_Jin_Zhou_Szolovits_2020}: includes a deletion-based selection mechanism that selects words with the most significant impact on the final decision outcome, aiming to preserve semantic similarity. A word replacement mechanism is designed to generate test cases through synonym extraction, lexical checking, and semantic similarity checking.

\subsection{Evaluation indicators}
We choose five evaluation indicators for the experiment:

(1) \emph{Success rate} (\emph{S-rate})~\cite{morris2020textattack2}, which indicates the proportion of test cases generated by the test method that are able to mislead the threat model out of all the tested examples. In this experiment, its formula can be expressed as follows:
\begin{equation}
\text{S-rate}=\frac{N_{suc}}{N}\times100\%
\end{equation}
where, $N_{suc}$ is the number of test cases that mislead threat models successfully, and $N$ is the total number of input examples ($N$ = 1,000 in our experiment) for the current test method.

(2) \emph{Change rate} (\emph{C-rate})~\cite{morris2020textattack2}, which represents the average proportion of the changed words in the original text. C-rate can be expressed as:
\begin{equation}
\text {C-rate }=\frac{1}{N_{suc}} \sum_{k=1}^{N_{suc}} \frac{\operatorname{diff} I_k}{\operatorname{len}\left(I_k\right)}\times100\%
\end{equation}
where $\operatorname{diff} I_k$ represents the number of words replaced in the input $I_k$ and $\operatorname{len}$($\cdot$) represents the sequence length. C-rate is an indicator designed to measure the difference in content between the generated test cases and the original examples.

(3) \emph{Perplexity} (\emph{PPL})~\cite{morris2020textattack2}, an indicator used to assess the fluency of textual test cases. Perplexity is defined as the exponentiated average negative log-likelihood of a sequence. If we have a tokenized sequence $I$=($w_1$,$w_2$,\dots,$w_n$), then the perplexity of $I$ is:
\begin{equation}
\operatorname{PPL}(I)=\exp \left\{-\frac{1}{n} \sum_i^n \log p_\theta\left(w_i \mid w_{<i}\right)\right\}
\end{equation}
where $\log p_\theta\left(w_i \mid w_{<i}\right)$ is the log-likelihood of the $i$-th token conditioned on the preceding tokens $w_{<i}$ according to the language model. Intuitively, given the language model for computing PPL, the more fluent the test case, the less confusing it is.

(4) \emph{Time overhead} (\emph{T-O})~\cite{morris2020textattack2}, which refers to the average time it takes for a test method to generate a successful test case.

(5) \emph{Query number} (\emph{Q-N})~\cite{morris2020textattack2}, which represents the number of times a test method needs to query the threat model on average for each successful test case generated. The query number and time overhead together reflect the efficiency of the test method.
\section{Experiment results and analyses}\label{sec5}
In this section, we present four research questions and discuss the experimental results.
\subsection{RQ1: How is the quality of the generated test cases by ABFS for different threat models and datasets?}
We compare the success rate, change rate, and PPL of the test cases generated by ABFS with those of baselines, and Table~\ref{tab1} shows the experimental results across five threat models and three datasets. From the perspective of identifying robustness flaws, PWWS achieves an average success rate of 68.177\% across all threat models, outperforming other baselines. On the other hand, we find that PWWS achieves a test success rate of up to 80.308\% on DNN-based NLP software~\cite{ren2019generating}; the success rates of other baselines also decrease to some extent, indicating the limited effectiveness of existing testing methods in LLM-based software testing. In most cases, ABFS achieves higher success rates across datasets from different domains and threat models with varying parameter sizes. In experiments using AG's News to test Llama2-13b, ABFS achieves a success rate of 70.772\%. In contrast, the baseline success rates are 25.403\%, 13.079\%, 45.955\%, 55.531\%, and 58.276\%, respectively, demonstrating the higher effectiveness of ABFS in identifying robustness flaws. Unlike traditional software, LLM-based software relies on the knowledge embedded in LLMs for decision-making rather than developer-defined system rules. Confidence in software quality can only be established by uncovering robustness flaws during testing. By breaking the limits of detectable robustness flaws, ABFS provides a deeper understanding of software behavior under adversarial conditions, which is crucial for ensuring reliability in deployment within high-risk environments. Although baselines like PWWS can achieve success rates close to ABFS in a few cases, the additional errors identified by ABFS reveal more subtle fault cases that are often overlooked. These cases expose minor weaknesses in the decision boundary of LLM-based software and identify inputs to which the software is particularly vulnerable, aiding in more precise robustness assessments before deployment.

The concept of adversarial test cases imposes high requirements on text naturalness. The change rate and PPL quantitatively measure the similarity and naturalness between test cases and original inputs, providing metrics that are more reproducible than human evaluation~\cite{10298415}. Thanks to its simple yet effective search method, ABFS achieves better similarity and naturalness under all experimental conditions. For the financial sentiment analysis dataset FP, ABFS achieves a change rate and PPL of only 0.914\% and 46.548, respectively. Test cases with low change rates pose a more significant practical threat, as they appear more like regular inputs yet lead the software to produce entirely incorrect decisions. These minor yet impactful modifications expose vulnerabilities that are difficult to detect in real-world deployment, helping identify potential attack vectors that attackers may exploit. High naturalness in test cases preserves semantic consistency with the original input, making the inputs still appear natural and understandable to humans. This is crucial during software testing because if test case modifications are too extreme, they may no longer represent real-world input patterns, diminishing the practical significance of the testing results. By generating natural test cases, ABFS better evaluates the software's tolerance to subtle perturbations.
\begin{table}
\centering
\caption{Comparison of the quality of test cases generated by the five baselines and ABFS.\\
\footnotesize Note: We mark the best result within a specific setting with \textbf{bold face}.
}
\label{tab1}
\setlength{\tabcolsep}{1pt} 
\footnotesize
\begin{tabular}{ccccccccc}
\hline
\textbf{Dataset}                                                                     & \textbf{Model}                                                                     & \textbf{Indicator} & \textbf{CheckList} & \textbf{StressTest} & \textbf{PWWS}   & \textbf{TextBugger} & \textbf{TextFooler} & \textbf{ABFS}   \\ \hline
\multirow{15}{*}{\textbf{FP}}                                                  & \multirow{3}{*}{\textbf{\begin{tabular}[c]{@{}c@{}}Mistral\\ -7b\end{tabular}}}    & \textbf{S-rate}    & 72.613             & 46.624              & 96.818          & 89.263              & 97.047              & \textbf{99.012} \\
                                                                               &                                                                                    & \textbf{C-rate}    & 1.863              & 8.556               & 1.182           & 2.013               & 1.195               & \textbf{0.914}  \\
                                                                               &                                                                                    & \textbf{PPL}       & 58.827             & 50.321              & 52.289          & 54.519              & 51.549              & \textbf{48.694} \\ \cmidrule(lr){2-9} 
                                                                               & \multirow{3}{*}{\textbf{\begin{tabular}[c]{@{}c@{}}Llama2\\ -13b\end{tabular}}}    & \textbf{S-rate}    & 32.447             & 34.676              & 84.676          & 72.442              & 79.449              & \textbf{95.654} \\
                                                                               &                                                                                    & \textbf{C-rate}    & 2.248              & 6.511               & 1.325           & 2.468               & 1.865               & \textbf{1.049}  \\
                                                                               &                                                                                    & \textbf{PPL}       & 59.444             & 54.862              & 54.653          & 58.001              & 54.903              & \textbf{49.471} \\ \cmidrule(lr){2-9} 
                                                                               & \multirow{3}{*}{\textbf{\begin{tabular}[c]{@{}c@{}}Internlm2\\ -20b\end{tabular}}} & \textbf{S-rate}    & 66.085             & 39.024              & 70.582          & 73.517              & 71.579              & \textbf{78.503} \\
                                                                               &                                                                                    & \textbf{C-rate}    & 1.945              & 9.423               & 1.167           & 3.004               & 1.239               & \textbf{0.938}  \\
                                                                               &                                                                                    & \textbf{PPL}       & 60.032             & 49.299              & 52.134          & 55.409              & 53.275              & \textbf{48.311} \\ \cmidrule(lr){2-9} 
                                                                               & \multirow{3}{*}{\textbf{\begin{tabular}[c]{@{}c@{}}Yi\\ -34b\end{tabular}}}        & \textbf{S-rate}    & 44.854             & 45.436              & 61.668          & 62.796              & 62.765              & \textbf{67.284} \\
                                                                               &                                                                                    & \textbf{C-rate}    & 1.908              & 7.95                & 1.473           & 2.087               & 1.619               & \textbf{0.968}  \\
                                                                               &                                                                                    & \textbf{PPL}       & 57.631             & 50.606              & 52.510          & 57.309              & 52.827              & \textbf{47.999} \\ \cmidrule(lr){2-9} 
                                                                               & \multirow{3}{*}{\textbf{\begin{tabular}[c]{@{}c@{}}Llama2\\ -70b\end{tabular}}}    & \textbf{S-rate}    & 37.772             & 46.866              & 83.232          & 80.953              & \textbf{88.912}     & 85.560          \\
                                                                               &                                                                                    & \textbf{C-rate}    & 2.349              & 8.738               & 2.089           & 2.122               & 2.176               & \textbf{0.989}  \\
                                                                               &                                                                                    & \textbf{PPL}       & 56.513             & 53.002              & 53.459          & 55.724              & 57.417              & \textbf{46.548} \\ \hline
\multirow{15}{*}{\textbf{\begin{tabular}[c]{@{}c@{}}AG's\\ News\end{tabular}}} & \multirow{3}{*}{\textbf{\begin{tabular}[c]{@{}c@{}}Mistral\\ -7b\end{tabular}}}    & \textbf{S-rate}    & 33.387             & 31.849              & \textbf{44.163} & 37.713              & 39.629              & 43.399          \\
                                                                               &                                                                                    & \textbf{C-rate}    & 4.512              & 9.281               & 3.804           & 3.856               & 2.796               & \textbf{0.771}  \\
                                                                               &                                                                                    & \textbf{PPL}       & 71.567             & 67.293              & 64.817          & 67.737              & 64.882              & \textbf{45.643} \\ \cmidrule(lr){2-9} 
                                                                               & \multirow{3}{*}{\textbf{\begin{tabular}[c]{@{}c@{}}Llama2\\ -13b\end{tabular}}}    & \textbf{S-rate}    & 25.403             & 13.079              & 45.955          & 55.531              & 58.276              & \textbf{70.772} \\
                                                                               &                                                                                    & \textbf{C-rate}    & 1.509              & 8.733               & 12.879          & 6.004               & 1.646               & \textbf{0.935}  \\
                                                                               &                                                                                    & \textbf{PPL}       & 60.183             & 54.108              & 56.222          & 56.559              & 54.813              & \textbf{43.816} \\ \cmidrule(lr){2-9} 
                                                                               & \multirow{3}{*}{\textbf{\begin{tabular}[c]{@{}c@{}}Internlm2\\ -20b\end{tabular}}} & \textbf{S-rate}    & 12.097             & 8.034               & 54.017          & 47.589              & 49.591              & \textbf{70.744} \\
                                                                               &                                                                                    & \textbf{C-rate}    & 1.862              & 8.387               & 3.271           & 9.783               & 3.194               & \textbf{1.316}  \\
                                                                               &                                                                                    & \textbf{PPL}       & 51.216             & 47.607              & 54.665          & 63.792              & 58.446              & \textbf{45.343} \\ \cmidrule(lr){2-9} 
                                                                               & \multirow{3}{*}{\textbf{\begin{tabular}[c]{@{}c@{}}Yi\\ -34b\end{tabular}}}        & \textbf{S-rate}    & 14.513             & 4.746               & 73.676          & 73.193              & 73.495              & \textbf{77.284} \\
                                                                               &                                                                                    & \textbf{C-rate}    & 1.532              & 6.968               & 1.451           & 4.398               & 2.529               & \textbf{0.796}  \\
                                                                               &                                                                                    & \textbf{PPL}       & 52.795             & 49.262              & 49.783          & 61.265              & 53.583              & \textbf{45.165} \\ \cmidrule(lr){2-9} 
                                                                               & \multirow{3}{*}{\textbf{\begin{tabular}[c]{@{}c@{}}Llama2\\ -70b\end{tabular}}}    & \textbf{S-rate}    & 13.018             & 9.591               & \textbf{56.527} & 33.564              & 38.363              & 55.219          \\
                                                                               &                                                                                    & \textbf{C-rate}    & 2.094              & 6.401               & 1.971           & 4.347               & 2.204               & \textbf{1.159}  \\
                                                                               &                                                                                    & \textbf{PPL}       & 54.652             & 53.808              & 53.367          & 54.701              & 49.639              & \textbf{44.575} \\ \hline
\multirow{15}{*}{\textbf{MR}}                                                  & \multirow{3}{*}{\textbf{\begin{tabular}[c]{@{}c@{}}Mistral\\ -7b\end{tabular}}}    & \textbf{S-rate}    & 45.507             & 46.302              & 49.503          & 47.508              & 48.505              & \textbf{51.556} \\
                                                                               &                                                                                    & \textbf{C-rate}    & 6.710              & 7.662               & 6.145           & 5.043               & 6.463               & \textbf{0.931}  \\
                                                                               &                                                                                    & \textbf{PPL}       & 88.198             & 80.945              & 78.205          & 73.389              & 71.028              & \textbf{54.098} \\ \cmidrule(lr){2-9} 
                                                                               & \multirow{3}{*}{\textbf{\begin{tabular}[c]{@{}c@{}}Llama2\\ -13b\end{tabular}}}    & \textbf{S-rate}    & 18.612             & 13.273              & 89.138          & 46.099              & 62.895              & \textbf{98.064} \\
                                                                               &                                                                                    & \textbf{C-rate}    & 1.944              & 12.935              & 1.124           & 4.146               & 1.964               & \textbf{1.047}  \\
                                                                               &                                                                                    & \textbf{PPL}       & 65.365             & 52.799              & 55.343          & 71.343              & 67.194              & \textbf{50.526} \\ \cmidrule(lr){2-9} 
                                                                               & \multirow{3}{*}{\textbf{\begin{tabular}[c]{@{}c@{}}Internlm2\\ -20b\end{tabular}}} & \textbf{S-rate}    & 36.574             & 12.568              & 81.512          & 82.514              & 83.586              & \textbf{83.790} \\
                                                                               &                                                                                    & \textbf{C-rate}    & 1.914              & 7.976               & 1.816           & 6.541               & 2.079               & \textbf{1.016}  \\
                                                                               &                                                                                    & \textbf{PPL}       & 67.764             & 62.167              & 63.426          & 70.614              & 63.907              & \textbf{58.598} \\ \cmidrule(lr){2-9} 
                                                                               & \multirow{3}{*}{\textbf{\begin{tabular}[c]{@{}c@{}}Yi\\ -34b\end{tabular}}}        & \textbf{S-rate}    & 21.795             & 17.465              & 80.239          & 82.951              & 79.964              & \textbf{85.361} \\
                                                                               &                                                                                    & \textbf{C-rate}    & 1.918              & 6.124               & 1.084           & 4.187               & 2.009               & \textbf{0.964}  \\
                                                                               &                                                                                    & \textbf{PPL}       & 68.956             & 62.005              & 61.203          & 75.764              & 66.749              & \textbf{56.913} \\ \cmidrule(lr){2-9} 
                                                                               & \multirow{3}{*}{\textbf{\begin{tabular}[c]{@{}c@{}}Llama2\\ -70b\end{tabular}}}    & \textbf{S-rate}    & 21.198             & 30.465              & 50.956          & 41.725              & 49.011              & \textbf{49.947} \\
                                                                               &                                                                                    & \textbf{C-rate}    & 2.647              & 7.707               & 2.131           & 3.944               & 2.653               & \textbf{1.344}  \\
                                                                               &                                                                                    & \textbf{PPL}       & 73.284             & 63.019              & 63.351          & 68.124              & 65.966              & \textbf{57.309} \\ \hline
\end{tabular}
 \vspace{-0.4cm}
\end{table}
\begin{mdframed}[backgroundcolor=gray!20, linecolor=black]
\textbf{Answer to RQ1:} 
ABFS demonstrates better test effectiveness, with an average success rate of 6.248\% higher than the optimal baseline PWWS. The generated test cases are more natural, with an average perplexity of 8.161 lower than the optimal baseline.
This indicates that ABFS can detect more realistic robustness flaws, aiding in a more comprehensive evaluation before software deployment.
\end{mdframed}
\subsection{RQ2: Can ABFS perform tests more efficiently?}
\begin{figure}[t]
 \centering
\includegraphics[width=1\linewidth]{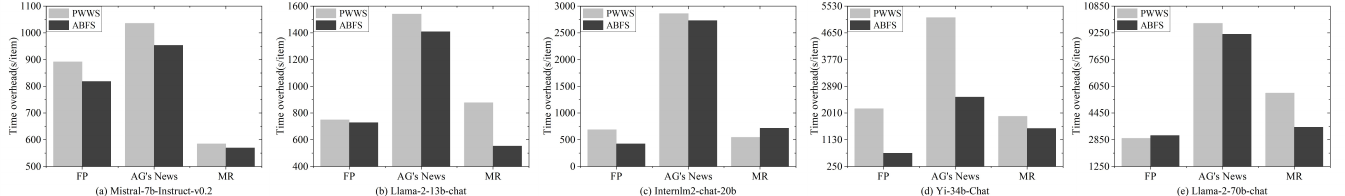}\\
 \caption{Results of test time overhead on different datasets and threat models (want $\downarrow$).} 
 \label{Fig5}
 \vspace{-0.3cm}
\end{figure}

\begin{figure}[t]
 \centering
\includegraphics[width=1\linewidth]{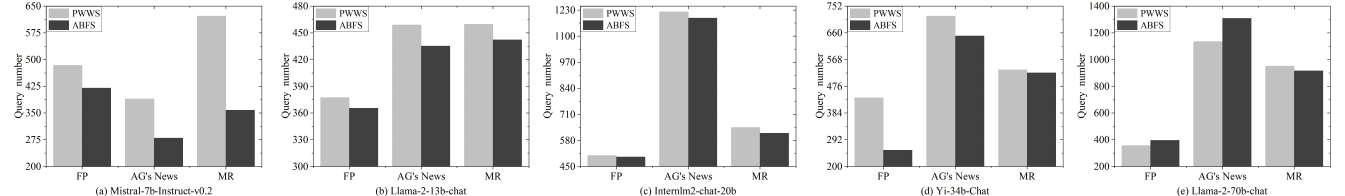}\\
 \caption{Results of test query number for different datasets and threat models (want $\downarrow$).} 
 \label{Fig6}
\end{figure}

In addition to test case quality, our study also focuses on the efficiency of testing methods, including time overhead and the number of queries. We use PWWS, the currently best-performing baseline shown in RQ1, to compare with ABFS across various datasets and threat models. Fig.~\ref{Fig5} presents the results for time overhead, indicating that ABFS requires less time to generate test cases; when testing Llama2-13b, ABFS is on average 159.009 seconds per item faster than PWWS. Fig.~\ref{Fig6} displays the results for the number of queries, showing that ABFS generates more successful test cases with fewer queries, requiring up to 264.217 fewer queries to the threat model per successful test case compared to PWWS. This efficiency is primarily due to the adaptive BFS, which maintains a priority queue to ensure that each queried node has the best perturbation effect in the entire queue, significantly reducing redundant queries. The adaptive control strategy dynamically adjusts the search path based on context and software feedback, reducing time overhead~\cite{ZHANG2020100647}. This efficiency is critical in LLM-based software testing, as it significantly lowers testing costs, such as reducing API call expenses on cloud platforms, and enables a deeper robustness evaluation within limited time constraints.
\vspace{0.1cm}
\begin{mdframed}[backgroundcolor=gray!20, linecolor=black]
\textbf{Answer to RQ2:} 
ABFS generates each successful test case an average of 584.565 seconds faster than the optimal baseline and requires fewer query attempts. This efficiency significantly reduces testing costs while enabling more in-depth robustness evaluations within limited time constraints.
\end{mdframed}
\subsection{RQ3: Does each of the methodological components proposed in this work improve the effectiveness and efficiency of ABFS?}
To evaluate the effectiveness of using WordNet to establish the transformation space, BFS-based test case search, and adaptive WIR calculation, we conduct an ablation study of ABFS on three datasets. Table~\ref{tab2} presents the robustness testing results on Llama2-13b using popular methods for establishing transformation spaces. ABFS with WordNet-based transformation space (ABFS) achieves higher testing effectiveness and generates more natural test cases. Although ABFS using random word insertion, deletion, and replacement to build the transformation space (ABFS-RW) achieves the fastest testing efficiency, its test success rate is on average 82.192\% lower than ABFS, making it inadequate for fully evaluating the robustness of LLM-based software. ABFS remains the most efficient testing method aside from ABFS-RW. The significant impact of ABFS on testing can be attributed to WordNet, a lexicon created by linguistic experts that provides more accurate and natural word substitutions, allowing test cases to retain the original text's semantics and structure to the greatest extent possible, thus achieving better testing results in robustness testing.
\begin{table}[]
\caption{The results of robustness testing on Llama2-13b using popular transformation space establishing methods, including random character insertion, deletion, and replacement (ABFS-RC); random word insertion, deletion, and replacement (ABFS-RW); synonym replacement with GloVe word embeddings (ABFS-G); and synonym replacement with WordNet (ABFS).
}
\label{tab2}
\begin{tabular}{ccccccc}
\hline
\textbf{Dataset} & \textbf{Method} & \textbf{S-rate} & \textbf{C-rate} & \textbf{PPL} & \textbf{T-O} & \textbf{Q-N} \\ \hline
\multirow{4}{*}{\textbf{FP}} & ABFS-RC & 58.893 & 1.157 & 54.752 & 1189.261 & 673.435 \\
 & ABFS-RW & 11.244 & 1.263 & 46.455 & \textbf{247.773} & \textbf{145.136} \\
 & ABFS-G & 93.442 & 1.104 & 51.736 & 1799.369 & 1940.360 \\
 & ABFS & \textbf{95.654} & \textbf{1.049} & \textbf{49.471} & 728.474 & 365.458 \\ \hline
\multirow{4}{*}{\textbf{AG's News}} & ABFS-RC & 66.125 & 1.012 & 46.993 & 1911.107 & 929.528 \\
 & ABFS-RW & 5.137 & 1.727 & 45.679 & \textbf{318.939} & \textbf{161.776} \\
 & ABFS-G & 68.236 & 0.981 & 44.041 & 5687.183 & 1122.451 \\
 & ABFS & \textbf{70.772} & \textbf{0.935} & \textbf{43.816} & 1408.859 & 435.11 \\ \hline
\multirow{4}{*}{\textbf{MR}} & ABFS-RC & 24.816 & 1.128 & 59.590 & 1851.186 & 915.025 \\
 & ABFS-RW & 1.533 & 1.117 & 47.305 & \textbf{240.548} & \textbf{74.305} \\
 & ABFS-G & 77.445 & 1.240 & 58.941 & 4425.04 & 2583.54 \\
 & ABFS & \textbf{98.064} & \textbf{1.047} & \textbf{50.526} & 554.014 & 442.180 \\ \hline
\end{tabular}
\end{table}

Table~\ref{tab3} demonstrates the effectiveness of ABFS's search strategy. We design an adaptive BFS specifically for the robustness testing of LLM-based NLP software; to our knowledge, no current research applies adaptive strategy and BFS in the robustness testing of LLM-based software. Regarding test effectiveness, the BFS-based testing method ($w/o$ Ada) achieves 7.815\%-22.633\% higher effectiveness than the standard greedy search testing method ($w/o$ BFS$\&$Ada). This shows that BFS enables a more comprehensive exploration of the transformation space, avoiding common local optima issues seen in greedy search. Moreover, ABFS with adaptive WIR calculation further improves testing effectiveness and significantly enhances testing efficiency. By dynamically adjusting each word's perturbation order based on historical information, adaptive WIR more accurately identifies words with significant impact on software output, optimizing the synonym replacement perturbation process. Therefore, ABFS can dynamically adapt to texts of varying lengths and domains, generating effective and natural test cases, which is critical for testing LLM-based NLP software.
\begin{table}[]
\caption{Ablation study results of standard greedy search (\textit{w/o} BFS\&Ada), adaptive greedy search (\textit{w/o} BFS), and standard BFS (\textit{w/o} Ada) on Llama2-13b.
}
\label{tab3}
\begin{tabular}{ccccccc}
\hline
\textbf{Dataset} & \textbf{Method} & \textbf{S-rate} & \textbf{C-rate} & \textbf{PPL} & \textbf{T-O} & \textbf{Q-N} \\ \hline
 & \textit{w/o} BFS\&Ada & 84.676 & 1.325 & 54.653 & 749.829 & 377.265 \\
 & \textit{w/o} BFS & 82.344 & 1.145 & 51.315 & 738.906 & 373.320 \\
 & \textit{w/o} Ada & 94.149 & 1.089 & 50.471 & 942.308 & 385.456 \\
\multirow{-4}{*}{\textbf{FP}} & ABFS & \textbf{95.654} & \textbf{1.049} & \textbf{49.471} & \textbf{728.474} & \textbf{365.458} \\ \hline
 & \textit{w/o} BFS\&Ada & 45.955 & 12.879 & 56.222 & 1540.594 & 458.881 \\
 & \textit{w/o} BFS & 51.116 & 0.952 & 45.355 & 1296.730 & 446.236 \\
 & \textit{w/o} Ada & 68.588 & 1.005 & 44.815 & 2125.702 & 553.264 \\
\multirow{-4}{*}{\textbf{AG's News}} & ABFS & \textbf{70.772} & \textbf{0.935} & \textbf{43.816} & \textbf{1408.859} & \textbf{435.11} \\ \hline
 & \textit{w/o} BFS\&Ada & 89.138 & 1.124 & 55.343 & 877.955 & 459.885 \\
 & \textit{w/o} BFS & 93.025 & 1.118 & 54.696 & 839.791 & 449.874 \\
 & \textit{w/o} Ada & 96.953 & 1.090 & 51.833 & 954.642 & 455.276 \\
\multirow{-4}{*}{\textbf{MR}} & ABFS & \textbf{98.064} & \textbf{1.047} & \textbf{50.526} & \textbf{554.014} & \textbf{442.180} \\ \hline
\end{tabular}
\end{table}
\begin{mdframed}[backgroundcolor=gray!20, linecolor=black]
\textbf{Answer to RQ3:} The ABFS, which integrates synonym replacement via WordNet, adaptive WIR, and BFS, demonstrates greater testing effectiveness and efficiency in the robustness testing of LLM-based NLP software.
\end{mdframed}
\subsection{RQ4: How transferable are the test cases generated by ABFS?}
We select the three baselines with the highest test success rates to evaluate the transferability of test cases generated by ABFS on each dataset. Table~\ref{tab4} presents the comparative results for test effectiveness, where ``13b→70b" indicates the success rate when test cases are transferred from Llama2-13b to Llama2-70b and vice versa. We find that, for different types of datasets and threat models with varying parameter sizes, the test cases generated by ABFS maintain higher success rates after transfer. For instance, on the AG's News dataset, ABFS achieves a transfer success rate of 82.418\% for test cases generated on Llama2-13b when transferred to Llama2-70b, while the next-best method, TextBugger, only achieves 57.732\%. ABFS conducts a global exploration of the transformation space through adaptive BFS, generating perturbations that impact the deep feature representations of LLMs, thereby uncovering more general robustness flaws in LLM-based software. Many advanced LLMs have built-in mechanisms to detect and filter out conspicuously abnormal inputs. LLM-based software often identifies and ignores test cases with low naturalness due to their prominent anomalous features. Using WordNet for synonym replacement, combined with a search strategy that ensures naturalness, ABFS generates test cases capable of bypassing these mechanisms to influence software outputs directly. This cross-model adaptability reduces the time and resource costs of generating test cases separately for each LLM, enhancing the broader applicability and long-term value of testing. Additionally, the success rate of transferring test cases from Llama2-70b to Llama2-13b is generally higher than the reverse, suggesting that a larger parameter size may enhance the robustness of LLMs with the same architecture.
\vspace{0.1cm}
\begin{table}[]
\caption{The success rates of transferred adversarial test cases on the three datasets (want $\uparrow$).\\
\footnotesize Note: We use 13b to represent Llama2-13b and 70b to represent Llama2-70b.
}
\label{tab4}
\begin{tabular}{cccccc}
\hline
\textbf{Dataset} & \textbf{\begin{tabular}[c]{@{}c@{}}Transfer\\ relation\end{tabular}} & \textbf{PWWS} & \textbf{TextBugger} & \textbf{TextFooler} & \textbf{ABFS} \\ \hline
\multirow{2}{*}{\textbf{FP}} & 13b→70b & 71.966 & \textbf{74.532} & 71.923 & 66.786 \\
 & 70b→13b & 82.119 & 81.393 & 82.143 & \textbf{84.951} \\ \hline
\multirow{2}{*}{\textbf{AG's News}} & 13b→70b & 56.164 & 57.732 & 47.252 & \textbf{82.418} \\
 & 70b→13b & 83.991 & 85.361 & 82.595 & \textbf{87.441} \\ \hline
\multirow{2}{*}{\textbf{MR}} & 13b→70b & 51.582 & 41.651 & 34.046 & \textbf{61.045} \\
 & 70b→13b & 75.378 & 76.104 & 72.833 & \textbf{76.765} \\ \hline
\end{tabular}
\end{table}
\begin{mdframed}[backgroundcolor=gray!20, linecolor=black]
\textbf{Answer to RQ4:} The test cases generated by ABFS exhibit higher transferability, helping to save the time and resources required to test each threat model individually. This efficiency enhances the broad applicability of the testing process and increases its long-term value.
\end{mdframed}
\section{Threats to validity}\label{sec6}
Our experimental results validate the effectiveness of ABFS. However, we acknowledge that this effectiveness is subject to certain threats.

\textbf{Internal validity}. The main threat to internal validity arises from the hyperparameter settings in our experiments, such as the number of synonyms retained when establishing the transformation space and the maximum query count. To mitigate these threats, we adopted specific strategies. For baseline methods, we strictly followed the original paper settings to ensure that our experiments are fair and reproducible. For ABFS, we applied the same values for $k_1$, $k_2$, and $\alpha _j$ across all experiments on each dataset and adjusted the maximum query count based on the responses from the threat model. To further minimize this threat, we meticulously reviewed our implementation and provided all materials and code for further examination.

\textbf{External validity}. External validity faces challenges regarding the datasets and the methods' applicability. Our experiments focus primarily on testing English-language LLM-based software, which may limit the generalizability of ABFS to other languages. However, adapting ABFS to other linguistic contexts requires only minor input adjustments. We tested ABFS on datasets from three specific domains and five threat models with varying parameter sizes, giving us confidence that ABFS will be effective across various NLP software applications. Although the focus of this paper is on validating ABFS performance in text classification tasks, the core strategies behind ABFS, such as BFS-based test case search and adaptive WIR calculation, can be extended to text generation tasks with simple modifications, for example, by using metrics like perplexity or likelihood scores to guide the optimization process instead of classification confidence.
\section{Related work}\label{sec7}
With the increasing prevalence of deep learning-driven intelligent software across various industries, researchers and developers are becoming acutely aware of the importance of thoroughly verifying software robustness~\cite{10.1145/3691620.3695024,9978989}. In the NLP field, LLMs' longer inference times and stronger robustness than DNNs make the manual construction of test cases inefficient and costly. Therefore, this paper primarily explores automated testing methods for NLP software. Based on the target threat model, related work can be categorized into two areas: robustness of DNN-based and LLM-based NLP software.

\subsection{Robustness of DNN-based NLP software}
Numerous methods have been proposed for testing DNN-based NLP software. Shen et al.~\cite{10.1145/3551349.3556953} proposed a metamorphic testing approach, QAQA, which leverages transformation relations, semantically guided search, and enhanced test oracles to reduce false positives when testing question-answering software. He et al.~\cite{9402091} introduced referentially transparent inputs, using a segment parser to extract phrases and a bag-of-words model to test machine translation tasks. Xiao et al.~\cite{10298415} proposed LEAP, which applies a Levy-flight enhanced particle swarm optimization algorithm to test text classification software efficiently. These black-box methods rely solely on the input-output information of the threat model without accessing internal knowledge of the DNN, allowing for simple modifications to the testing process to be applied in LLM-based NLP software testing.

\subsection{Robustness of LLM-based NLP software}
Most research on LLM-based software testing has focused on prompt-oriented robustness testing.
Zhang et al.~\cite{10.1145/3691620.3695001} explored exploitable design characteristics in LLM-based software, generating trigger sequences from original texts to create toxic prompts using position-insensitive generation techniques. In response, Liu et al.~\cite{10.1145/3691620.3695018} leveraged LLMs to create toxic concept prompts, forming feature vectors using word embeddings and employing multilayer perceptrons to detect toxic prompts. Notably, most existing research focuses on jailbreaking threats. Pedro et al.~\cite{p2sql} automatically generated new malicious prompts by fine-tuning Mistral-7b on an initial set of handcrafted prompts, uncovering the P$_2$SQL vulnerability in five LLM-based software applications. Deng et al.~\cite{deng2024masterkey} trained a specialized LLM using a jailbreaking prompt dataset and applied reward strategies to enhance the model's ability to generate jailbreak prompts. However, jailbreaking attacks aim to induce LLMs to violate their original design constraints by constructing specific prompts, leading them to perform prohibited actions such as generating unethical content. In contrast, our work aims to test the ability of LLM-based software to make correct decisions under minor, imperceptible perturbations. This fundamental difference in testing goals and impact scope sets our approach apart.

Current robustness testing for example-oriented tasks predominantly uses manually or automatically constructed offline datasets. Yuan et al.~\cite{10.5555/3666122.3668671} proposed the BOSS benchmark for evaluating out-of-distribution robustness, covering five NLP tasks and twenty datasets. Dong et al.~\cite{10.1007/978-3-031-44693-1_53} considered the robustness of LLM-based software against real-world noise, constructing Noise-LLM, a dataset comprising five single perturbations and four mixed perturbations. Liu et al.~\cite{liu2023robustness} used existing DNN-based NLP software testing methods to generate test cases for contextual learning tasks, evaluating the robustness of LLMs over time. However, Zverev et al.~\cite{zverev2024can} found that LLMs struggle to decouple prompts from examples, and the complex interaction between these elements and LLMs implies that testing only prompts or examples does not fully reveal robustness flaws in LLM-based software. Inspired by Liu et al.~\cite{liu2023robustness}, ABFS adopts DNN-based NLP software testing methods as baselines, testing overall software input to provide a novel perspective and tool for studying robustness in LLM-based software. We refrain from using offline datasets as this static testing paradigm risks data leakage, which compromises fairness, and the rapid iteration of LLMs can make static data quickly outdated, reducing test accuracy.
ABFS automatically constructs real-time data by querying the threat model, ensuring better test coverage and flexibility than static testing.
\section{Conclusion}\label{sec8}
In this paper, we propose ABFS, an automated testing method designed to generate effective and natural adversarial test cases for LLM-based NLP software. ABFS treats the input prompt and example as a unified whole, introducing BFS to explore successful test cases in the high-dimensional text space, and further enhancing test case naturalness through an adaptive control strategy. We evaluate ABFS using three datasets, five threat models, and five baselines. The results show that ABFS achieves an average test effectiveness of 74.425\%, compared to the currently best-performing PWWS at 68.177\%. Compared to the baselines, the test cases generated by ABFS are consistently more natural, 
introducing minimal perturbations to the original input.
These test cases can simulate actual user inputs more realistically, enabling a more comprehensive and accurate assessment of software robustness. In future work, we will explore more effective methods to test the robustness of LLM-based software, aiming to improve both the quality of test cases and testing efficiency. We will also conduct evaluations across various downstream tasks and datasets.
\backmatter

\bibliography{sn-bibliography}

\end{document}